\documentclass[conference,10pt]{IEEEtran}
\IEEEoverridecommandlockouts

\usepackage{textcomp}
\usepackage{gensymb}
\usepackage{multirow}
\usepackage[ruled,linesnumbered,noend]{algorithm2e}

\usepackage{cite}
\usepackage{subcaption}
\usepackage{amsmath,amssymb,amsfonts}
\usepackage[ruled,linesnumbered,noend]{algorithm2e} 
\usepackage{graphicx}
\usepackage{textcomp}

\usepackage{tabularx}
\usepackage{makecell}

\usepackage{mathtools}
\DeclarePairedDelimiter{\ceil}{\lceil}{\rceil}

\usepackage{url}
\def\BibTeX{{\rm B\kern-.05em{\sc i\kern-.025em b}\kern-.08em
    T\kern-.1667em\lower.7ex\hbox{E}\kern-.125emX}}

\begin{document}

\title{An Analytical Model for Efficient Indoor THz Access Point Deployment
\thanks{This project was funded by CMU Portugal Program: CMU/TMP/0013/2017- THz Communication for Beyond 5G Ultra-fast Networks.}
}

\author{\IEEEauthorblockN{Rohit Singh \textsuperscript{1}, Douglas Sicker \textsuperscript{1,2}}
\IEEEauthorblockA{ \textsuperscript{1}  \textit{Engineering \& Public Policy}, \textit{Carnegie Mellon University}, Pittsburgh, USA \\
			       \textsuperscript{2}  \textit{School of Computer Science}, \textit{Carnegie Mellon University}, Pittsburgh, USA \\
Email: rohits1@andrew.cmu.edu, sicker@cmu.edu}
}

\maketitle

% The abstract is a short summary of the work to be presented in the article.
\section*{ABSTRACT}

Ultra-densification of user equipment (UE) and access points (APs) are anticipated to take a toll on the future spectrum needs. Higher frequency bands, such as mmWave ($30$-$300GHz$) and THz spectrum ($0.3$-$10THz$), can be used to cater to the high-throughput needs of ultra-dense networks. These high-frequency bands have a tremendous amount of \textit{green-filed contiguous spectrum}, ranging in hundreds of $GHz$. However, these bands, especially the THz bands, face numerous challenges, such as high spreading, absorption, and penetration losses. To combat these challenges, the THz-APs need to be either equipped with high transmit power, high antenna gains (i.e., narrow antenna beams), or limit the communication to short-ranges. All of these factors are bounded due to technical or economic challenges, which will result in a \textit{``distance-power dilemma"} while deciding on the deployment strategy of THz-APs. In this paper, we present an analytical model to deploy THz-APs in an indoor setting efficiently. We further show through extensive numerical analysis, the optimal number of APs and optimal room length for different blocks of the THz spectrum. Furthermore, these THz-APs need to be efficiently packed to avoid outages due to handoffs, which can add more complexity to the dilemma. To mitigate the packing problem, we propose two solutions over the optimal solution: (a) Radius Increase, and (b) Repeater Assistance, and present an analytical model for each.   

\section*{Keywords} Terahertz (THz), Indoor AP Placement, Energy Efficient, Spectrum Efficient, AP Fitting, Lambert's W-function

%%%%%%%%%%%%%%%%%%%%%%%%%%%%%%%%%%%%%%%%%%%%%%%%%%%%%%%%%%%%%
%%%%%%%%%%%%%%%%%%%%%%%%%%%%%%%%%%%%%%%%%%%%%%%%%%%%%%%%%%%%%
%%%%%%%%%%%%%%%%%%%%%%%%%%%%%%%%%%%%%%%%%%%%%%%%%%%%%%%%%%%%%

\setlength\belowcaptionskip{-0.2 in}
\setlength{\abovecaptionskip}{0.2 in}
\begin{figure*}[t]
\centering
\begin{subfigure}[]{3.5 in}
\includegraphics[width=3.8 in,height=1.7 in]{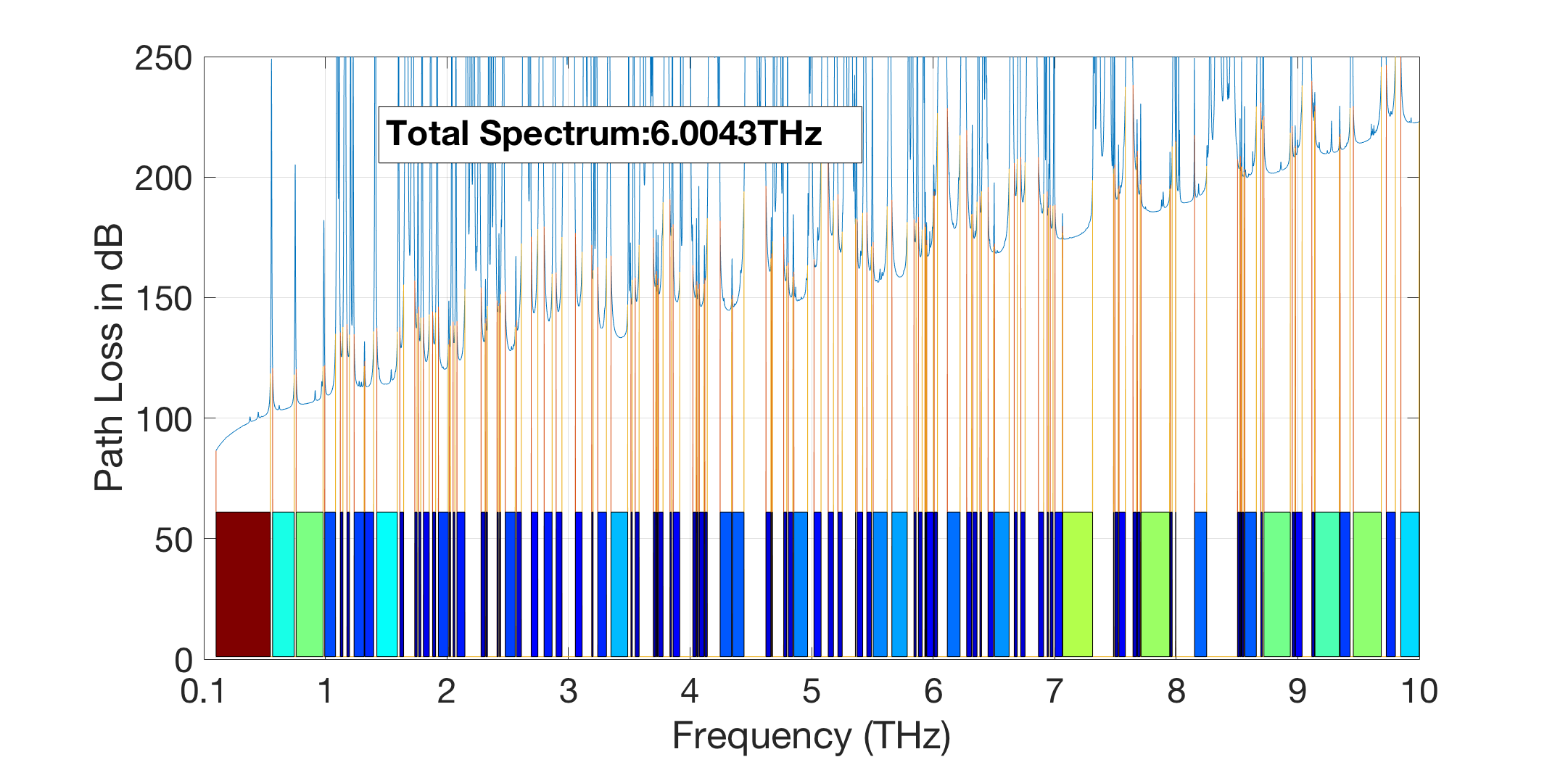}
\caption{Distance $=5m$.}
\label{La5m}
\end{subfigure}
~
\begin{subfigure}[]{3.5 in}
\includegraphics[width=3.5 in,height=1.7 in]{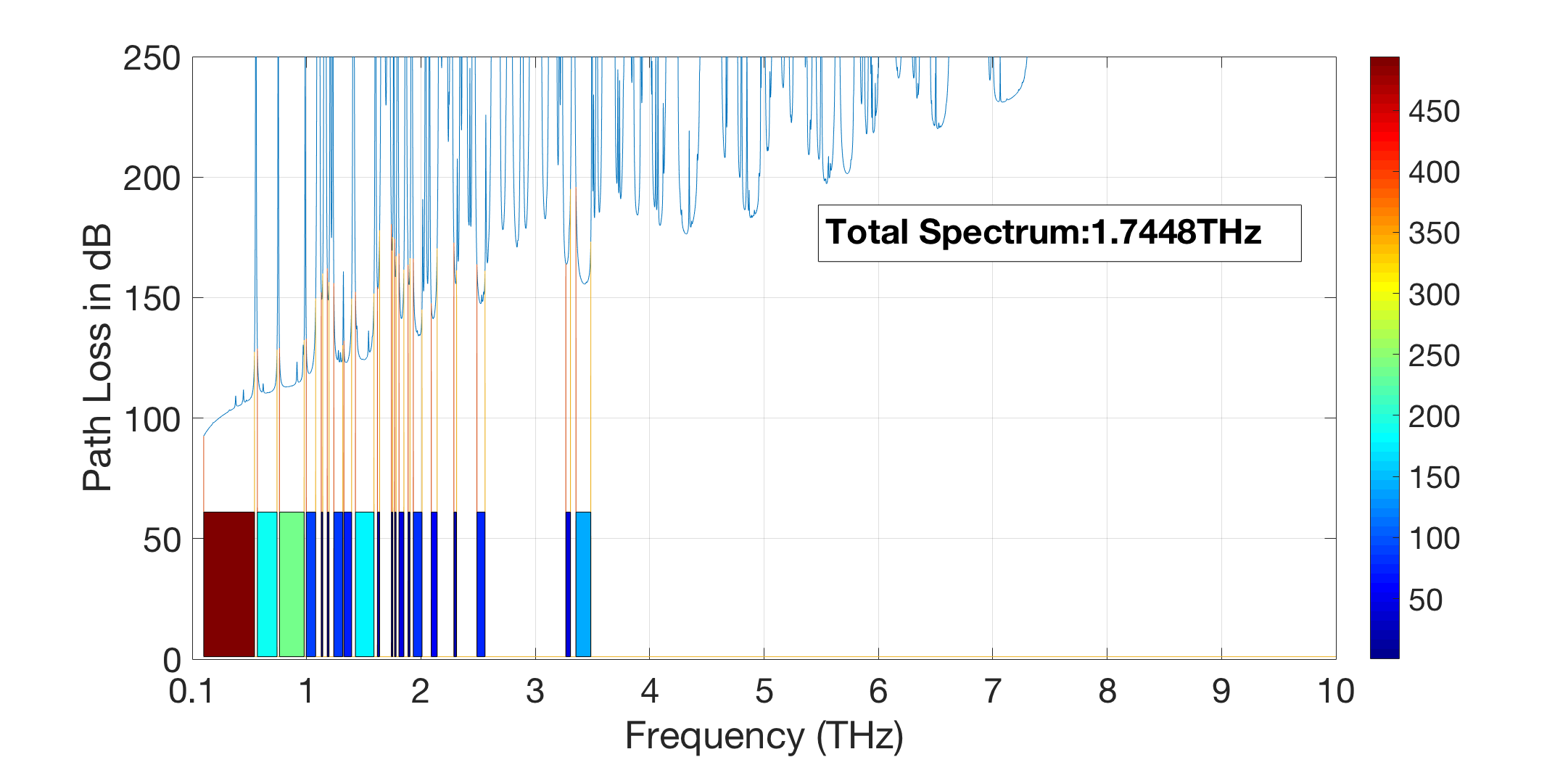}
\caption{Distance $=10m$.}
\label{La10m}
\end{subfigure}

\caption{THz spectrum sensitivity analysis for fixed relative humidity $\rho=60\%$ and loss cut-off value of $3dB$. The red and the orange lines define the start and the end frequencies of the sub-channel, and the color defines their contiguous bandwidths.}
\label{Loss}
\end{figure*}

\setlength{\intextsep}{0pt}% Remove \textfloatsep
\setlength\belowcaptionskip{9pt}
\setlength\abovecaptionskip{0pt}
\begin{table*}[t]  
\caption{A list of prospective sub-channels in each THz block and its charecteristics. Corresponding transmit power required by a single THz-AP to achieve $0.1Gbps/GHz$ spectral efficiency with antenna beamwidth $\delta$, $\rho=60\%$, and distance $5m$.}
\centering
\begin{tabular}{ |p{2cm}|p{1.5cm}|p{1cm}|p{1cm}|p{1cm}|p{1cm}|p{1cm}|p{1cm}|p{1cm}|p{1cm}|p{1cm}|p{1cm}|} 
\hline
\multicolumn{2}{|c|}{}	 
								& $f_1$ & $f_2$ & $f_3$ & $f_4$ & $f_5$ & $f_6$ & $f_7$ & $f_8$ & $f_9$ & $f_{10}$ \\ \hline
\multicolumn{2}{|c|}{\textbf{Center Frequencies $f_c$ in $THz$}}	
								&  $0.32$ & $1.51$ & $2.52$ & $3.42$ & $4.91$ & $5.72$ & $6.57$ & $7.19$ & $8.83$ & $9.57$ \\ \hline
\multicolumn{2}{|c|}{\textbf{Contiguous Bandwidth in $GHz$}}	 
								& $447$ & $169$ & $82$ & $137$ & $113$ & $126$ & $120$ & $246$ & $217$ & $230$ \\ \hline
											
\multirow{2}{*}{\makecell{\textbf{Required Power} \\ \textbf{in dB}}}
				     & $\delta=20\degree$ 	&  $-50.73$&$-33.42$&$-19.45$&$-14.23$&$2.06$&$10.85$&$22.84$&$	28.12$&$54.52$&$67.56$  \\ \cline{2-12}
				     & $\delta=10\degree$         &  $-62.77$&$-45.46$&$-31.49$&$-26.27$&$-9.98$&$-1.19$&$10.80$&$16.08$&$42.48$&$55.52$  \\ \hline
				     
\end{tabular}
\label{Tab1}
\vspace{-4mm}
\end{table*}

\section{Introduction}

A report by the International Telecommunication Union-Recommendation (ITU-R) predicts by the year 2030, the number of global mobile subscriptions will reach nearly $17Billion$, on which live and on-demand video will account for a significant portion of the traffic \cite{ITUR}. Although it is hard to predict $20$ years out concerning the specific user demands and needs, it is evident that to accommodate such ultra-dense networks, more spectrum will be required. Many applications, such as live streaming, on-demand video, online gaming, virtual reality (VR) and augmented reality (AR), are dependent on the availability of low-latency and high-throughput wireless access. Obtaining more spectrum in the lower and mid bands will be beneficial for this purpose. However, in the US, the mid-bands are highly occupied, and repurposing of the spectrum will take time, which can hinder the progress of technology. Moving to higher frequency bands, like THz spectrum ($0.1THz-10THz$), can help facilitate the advancement of technology and satisfy the growing demand.  

As we move higher in the frequency, losses such as spreading with distance, absorption due to water vapor, and penetration through objects, increases, which can limit the effective communication to short ranges, i.e., $<10m$ \cite{NextFront}. THz spectrum is sandwiched between the microwave and infrared bands, which makes it both beneficial and challenging for deployment \cite{NextFront, OurTPRC, IMECSur}. On the one hand, THz is not so sensitive to particles and objects compared to infrared bands, while, on the other hand, it has more \textit{green-filed contiguous spectrum} compared to mmWave bands. THz frequency cannot be extensively used in outdoor settings like traditional RF bands, except for some unique use-cases \cite{OurCCNC2}. However, THz-APs can be deployed indoors with dense coverage of users requiring ultra-low-latency and ultra-high-throughput.

%%%%%%%%%%%%%%%%%%%%%%%%%%%%%%%%%%%%%%%%%%%%%%%%%%%%%%%%%%%%%%%%%%%%%%%%%%%%%%%%%%%%%%%%%%%%%%%%%%%%%%%%%%%%%%%%%%%%%%%%%%
\subsection{The Dilemma}

Although THz has an abundance of the spectrum, nearly $~9.9THz$, which is more than what we can ever achieve in the lower RF bands, the total \textit{usable spectrum} is dependent on environmental parameters and specific use cases. THz frequencies are sensitive to molecular absorption, like $O_2$ and $H_2O$, which results in a tremendous amount of path loss. As shown in Fig. \ref{Loss}, the total usable spectrum diminishes from $6THz$ (Fig. \ref{La5m}) to $1.7THz$ (Fig. \ref{La10m}) when we increase the separation from $5m$ to $10m$. The sensitivity analysis shown in Fig. \ref{Loss} is dependent on the environment and the transceiver properties. In case of high humid scenarios ($\rho=100\%$ when humans exhale) or poor hardware (which cannot cut-off such narrow spikes), the usable spectrum will reduce drastically. Thus, there is a need to use this spectrum more efficiently. 

Based on our sub-channel analysis shown in Fig. \ref{Loss}, we divide the THz spectrum into $10$ bands, which are $1THz$ apart and identify as prospective carrier frequencies with the maximum available contiguous sub-channel bandwidth, as shown in Table \ref{Tab1}. For a $5m$ separation, almost every sub-channel has more than $100GHz$ of contiguous spectrum. The largest sub-channels are mostly at the lower and the upper ends of the spectrum. Although relatively more bandwidth is available at the upper end of the spectrum, the transmit power required to obtain as low as $0.1Gbps/GHz$ of spectral efficiency is enormous. Currently, THz hardware is still in the development phase and produces a low peak output power of nearly $0dBm$ \cite{HWT}. THz transceivers currently have a low breakdown power, high sensitivity to noise, and high manufacturing costs \cite{OurTPRC}. The problem will become more acute when we move to the higher frequency sub-channels. Thus, other than spectral efficiency there is also a need for THz-APs to be power efficient.

A solution to improve efficiency, as highlighted in Table \ref{Tab1}, is to improve the antenna gain (i.e., narrower antenna beamwidth $\delta$). Although narrower antenna beamwidth might help reduce the required transmit power \cite{OurGC}, it can also increase outages due to small-scale mobility \cite{OurCCNC1}, resulting in lower average throughput. Nevertheless, the use of narrow laser beams will also require these antennas to operate at lower transmit power for human safety concerns \cite{OurTPRC}. 

Another solution is to limit the communication to short-range $1$-$5m$. Shorter distance will not only reduce the need for a high transmit power, but also allow a more usable spectrum, as shown in Fig.\ref{Loss}. However, to ensure coverage, a room will need to be blanketed with THz-APs. This dense deployment of THz-APs will result in (a) resource wastage, (b) excessive handoffs, (c) increased backhaul cost, (d) increased cost of device and deployment, (e) increased human safety concerns (THz is thermal in nature), and (f) indoor aesthetic concerns. 

THz spectrum has multiple challenges, out of which ``Distance" and ``Transmit Power" can become the critical bottlenecks for indoor THz deployment \cite{DistBW} \cite{ComDP}. Approaches such as energy harvesting for nano-networks \cite{OptiEng} and efficient antenna design and strategies \cite{EDsg, FDsg2, HMAntAr, InfoShower, LastMile} have been explored in the THz domain. However, very few researchers have looked at how these THz-AP should be deployed based on room shape, room demand profile, and other environmental factors. The question is, for a given room shape and environment, whether to reduce distance, or reduce transmit power, or both, such that the spectral demand for the room is achieved.

%%%%%%%%%%%%%%%%%%%%%%%%%%%%%%%%%%%%%%%%%%%%%%%%%%%%%%%%%%%%%%%%%%%%%%%%%%%%%%%%%%%%%%%%%%%%%%%%%%%%%%%%%%%%%%%%%%%%%%%%%%

\subsection{Our Contribution}
\begin{itemize}
\item To solve the issue of "Distance-Power Dilemma" in the THz spectrum, in section \ref{Ana}, we propose an analytical model to quantify: \textit{how many equally powered THz-APs can be closely packed in any room?} 
\item We propose our analysis for a hypothetical linear room and extend it for a more general room with different shapes and sizes. We consider both homogeneous and heterogeneous user distribution within the room, which accounts for the number of users and its mobility).   
\item For energy efficiency, we consider an upper bound over the cumulative power used by all the $N$ optimal THz-APs to $\mathcal{P}_o$. To ensure a fair share of resources, each THz-AP is given an equal portion of the total power, i.e., $\mathcal{P}_o/N$. With each additional THz-AP, both the power and the coverage area decreases, which ensures an optimal number of THz-APs that can fit in the room. This constrained power model can also be considered for user safety, where the total power of the room can be constrained to $\mathcal{P}_o$.
\item Most of our analysis represents a widely studied form known as the Lambert's function \cite{Lambert}. $\mathcal{W}(\cdot)$ is the Lambert's W-function given by the inverse of $f(\mathcal{W}) = \mathcal{W}e^\mathcal{W}$.
\item In Section \ref{CPack}, we propose two analytical models to improve cell packing without requiring the need to increase the number of THz-APs, i.e., (a) Radius Increase and (b) Repeater Assistance
\item In section \ref{NuAna}, we identify $10$ potential sub-channels that can be used for future deployment. We provide an extensive numerical analysis of our analytical model for these channels. 
\end{itemize}

\setlength\belowcaptionskip{-0.2in}
\setlength{\abovecaptionskip}{0 in}
\begin{figure}[t]
\centering
\includegraphics[width=3.5 in,height=2 in]{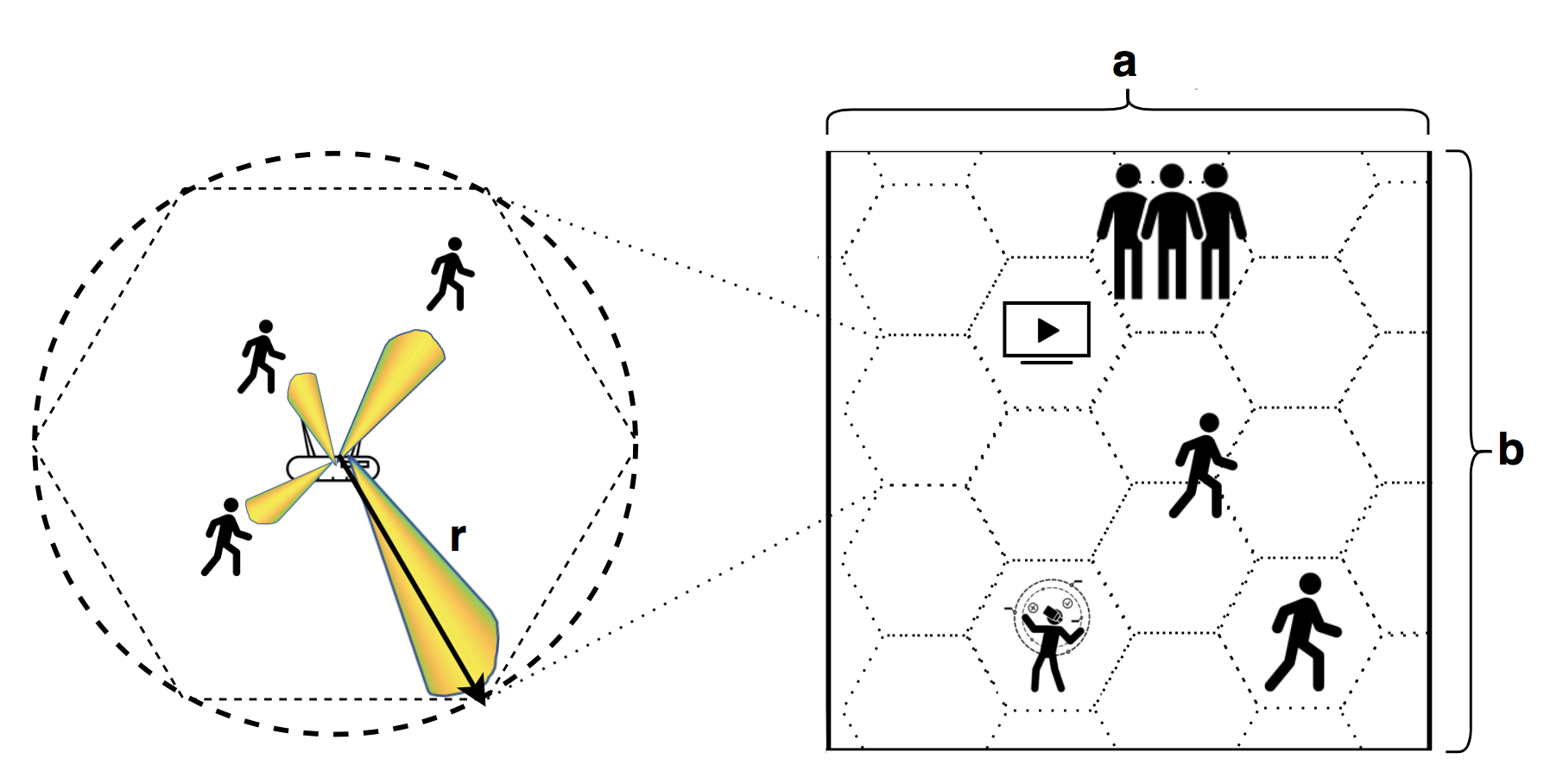}
\caption{THz Cell architecture for a rectangular room $aXb$.}
\label{CellArc}
\end{figure}

%%%%%%%%%%%%%%%%%%%%%%%%%%%%%%%%%%%%%%%%%%%%%%%%%%%%%%%%%%%%%%%%%%%%%%%%%%%%%%%%%%%%%%%%%%%%%%%%%%%%%%%%%%%%%%%%%%%%%%%%%%
%%%%%%%%%%%%%%%%%%%%%%%%%%%%%%%%%%%%%%%%%%%%%%%%%%%%%%%%%%%%%%%%%%%%%%%%%%%%%%%%%%%%%%%%%%%%%%%%%%%%%%%%%%%%%%%%%%%%%%%%%%
%%%%%%%%%%%%%%%%%%%%%%%%%%%%%%%%%%%%%%%%%%%%%%%%%%%%%%%%%%%%% %%%%%%%%%%%%%%%%%%%%%%%%%%%%%%%%%%%%%%%%%%%%%%%%%%%%%%%%%%%%%   
   
\section{THz-AP Analysis} \label{Ana}

In this section, we present our analytical model to find the optimal number of THz-APs required to fit any type of room.  
    
\subsection{System Model}

In this section, we briefly discuss the system models and assumptions that are applied in this paper. The path loss at the lower RF frequencies were mostly dominated by the spreading loss $\mathcal{L}_S$, which followed a simple free-space path loss (FSPL) model, i.e., $\mathcal{L}_S\propto f_c^2d^2$, where $f_c$ is the carrier frequency, and $d$ is the Euclidian distance. As we move to THz frequencies, the wavelengths get smaller and are more sensitive to water, oxygen molecules, and even some small particles. The transmittance of these waves can be obtained from Beer Lambert Law. Thus, the absorption loss $\mathcal{L}_A=e^{\mathcal{K}(f_c,\rho,T)d}$, where $\mathcal{K}$ is the medium absorption coefficient dependent on the frequency range, relative humidity $\rho$, and room temperature $T$, which can be calculated from the HIgh resolution TRANsmission molecular absorption (HITRAN) database as shown in \cite{ChaModelTHz}. Thus, the total path loss $\mathcal{L}_T$ is shown in Equation \ref{LossEq}.

\abovedisplayskip=-4pt
\belowdisplayskip=4pt
\begin{eqnarray}
 \mathcal{L}_T=\mathcal{L}_S(f_c,d)*\mathcal{L}_A (f_c,d,\rho)=(\frac{4 \pi d f_c}{c})^2 * e^{\mathcal{K}(f_c,\rho,T)d} 
\label{LossEq}
\end{eqnarray}

To find the spectral efficiency $\mathcal{S}=\mathcal{R}/\mathcal{B}$,where $\mathcal{R}$ is the requested data rate and $\mathcal{B}$ is the available bandwidth (dependent on the $f_c$,$\rho$, and $d$ as shown in Fig. \ref{Loss}), we use Shannon's Capacity theorem. The spectral efficiency at a given snapshot for a room with $J$ users is defined in Equation \ref{RateEq}, where $\mathcal{P}_t$ is the transmit power of the THz-AP, $G_t$ and $G_r$ are the transmitter and receiver gains dependent on the antenna beamwidth $\delta$, $N_f$ is the the noise power spectral density in $dB/Hz$,  and $c$ is the speed of light \cite{OurGC}. Since THz has very low transmit power, we can avoid intercell interference.

\abovedisplayskip=-4pt
\belowdisplayskip=4pt
\begin{eqnarray}
\mathcal{S}=\log_2(1+\frac{\mathcal{P}_t*G_t (\delta)*G_r (\delta)}{N_f*B_j*(4 \pi d f_c)/c)^2*e^{\mathcal{K}(f_c,\rho,T)d}})
\label{RateEq}
\end{eqnarray}

Let us consider a room of length $a$ and width $b$ with homogeneous distribution of $J$ different users. To ensure coverage the room has a blanket of THz-cells, a hexagonal cell, as shown in Fig \ref{CellArc}. We envision the THz-AP is at the center of the THz-cell, which uses narrow beams to dynamically beam steer based on user position. Without loss of generality, the beam steering process resembles a circular shape of radius $r$ or the coverage area of the THz-cell. The maximum Euclidian separation between the TX-RX within a THz-cell can be $r$.

%%%%%%%%%%%%%%%%%%%%%%%%%%%%%%%%%%%%%%%%%%%%%%%%%%%%%%%%%%%%%%%%%%%%%%%%%%%%%%%%%%%%%%%%%%%%%%%%%%%%%%%%%%%%%%%%%%%%%%%%%%

\subsection{Optimal Number of APs}

\begin{figure}[h]
\centering
\begin{subfigure}[]{1.5 in}
\includegraphics[width=1.5 in,height=1.7 in]{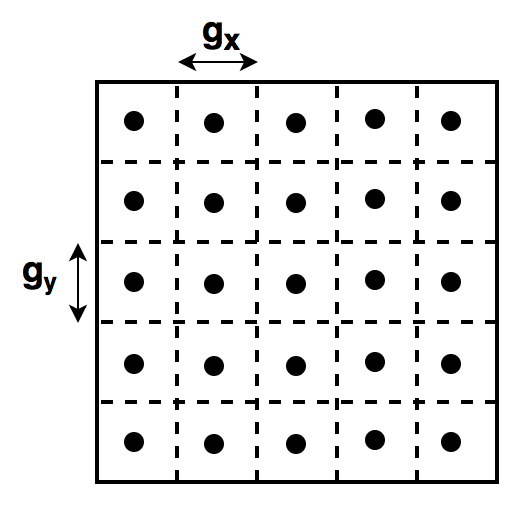}
\caption{A room with N Cells.}
\label{Rmod1}
\end{subfigure}
~
\begin{subfigure}[]{1.5 in}
\includegraphics[width=1.5 in,height=1 in]{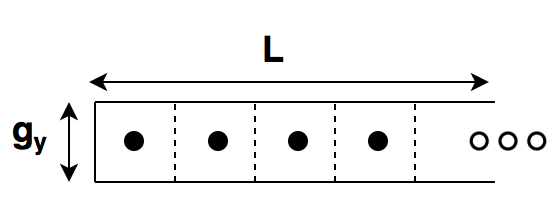}
\caption{A Hypothetical room.}
\label{Rmod2}
\end{subfigure}

\caption{A top view of a room with $N$ optimal THz-APs.}
\label{Rmod}
\end{figure}

To find the optimal number of THz-APs for a room we constrain the maximum transmit power of the cell to $\mathcal{P}_t=\mathcal{P}_o/N$, where $\mathcal{P}_o$ is the total power for the room and $N$ is the optimal number of THz-APs. Let us consider a homogeneous room that is divided into $N$ cells each of length $g_x$ and width $g_y$, as shown in Fig. \ref{Rmod1}. The total number of cells in the room is $N=N_xN_y$, where $N_x=a/g_x$ and $N_y=b/g_y$ are the number of cells on the x-axis and y-axis of the room respectively, where $g_x=g_y=2r$. Since each cell dimension is dependent on the assigned power, for each additional AP on the x-axis, the power for each cell will decrease and so will the radius $r$. Thus, the variables $N_x$ and $N_y$ form a recursive relation to each other and is hard to solve in this form. 

\setlength\belowcaptionskip{0 in}
\setlength{\abovecaptionskip}{0 in}

Let us dismantle the $N$ cell room and arrange it in a linear hypothetical room of length $L$ and width $g_y$, as shown in Fig. \ref{Rmod2}. This hypothetical room constraints the width of the room $b=g_y$. Without loss of generality we can derive the length of this hypothetical room as shown in Equation \ref{RomL}. 

\abovedisplayskip=0pt
\belowdisplayskip=4pt
\begin{eqnarray}
L=g_xN_xN_y=2rN
\label{RomL}
\end{eqnarray}

Using Equations \ref{RateEq}, \ref{RomL} and the power constraint of $\mathcal{P}_t=\mathcal{P}_o/N$ we can derive Equation \ref{ShanN}

\abovedisplayskip=0pt
\belowdisplayskip=4pt
\begin{eqnarray}
\begin{multlined}
2^\mathcal{S}-1=\frac{\frac{\mathcal{P}_o}{N}G_t (\delta)G_r (\delta)}{N_f(\frac{4\pi f_cr}{c})^2e^{\mathcal{K}r}} \\ 
						= \frac{\mathcal{P}_oG_t (\delta)G_r (\delta)c^2}{N*N_f(4\pi f_c)^2(\frac{L}{2N})^2e^{\frac{\mathcal{K}L}{2N}}}
\end{multlined}
\label{ShanN}
\end{eqnarray}

With simple manipulation on Equation \ref{ShanN} we have Equation \ref{OptN}, which is a function of $N$ and some constants, where $K_1=\frac{4\mathcal{P}_o G_t (\delta)*G_r (\delta) c^2}{N_f*(2^\mathcal{S}-1)(4\pi f_cL)^2}$ and $\tau_1=\mathcal{K}(f_c,\rho,T)L/2$. 

\abovedisplayskip=0pt
\belowdisplayskip=4pt
\begin{eqnarray}
 1/N e^{\tau_1/N}=K_1
\label{OptN}
\end{eqnarray}

Equation \ref{OptN} represents a widely studied form known as the Lambert's function \cite{Lambert}. Using this inverse form we can solve $N$ as shown in Equation \ref{OptiNval}, where $\mathcal{W}(\cdot)$ is the Lambert's W-function given by the inverse of $f(\mathcal{W}) = \mathcal{W}e^\mathcal{W}$. Since fractional number of THz-AP doesn't make sense, we use a ceiling function $\ceil*{\cdot}$ to round it off to the next integer value. 

\abovedisplayskip=0pt
\belowdisplayskip=4pt
\begin{eqnarray}
 N = \ceil*{\frac{\tau_1}{\mathcal{W}(\tau_1 K_1)}}
\label{OptiNval}
\end{eqnarray}

Thus, from Equation \ref{OptiNval} we can figure out the optimal number of THz-APs required for a room such that each cell has a spectral efficiency of $\mathcal{S}$, a total power consumption of $\mathcal{P}_o$, and THz-cell radius as shown in Equation \ref{OptiRval}.

 \abovedisplayskip=0pt
\belowdisplayskip=4pt
\begin{eqnarray}
 r = \ceil*{\frac{L\mathcal{W}(\tau_1 K_1)}{2\tau_1}}
\label{OptiRval}
\end{eqnarray}
%%%%%%%%%%%%%%%%%%%%%%%%%%%%%%%%%%%%%%%%%%%%%%%%%%%%%%%%%%%%%%%%%%%%%%%%%%%%%%%%%%%%%%%%%%%%%%%%%%%%%%%%%%%%%%%%%%%%%%%%%%

\setlength\belowcaptionskip{-0.1 in}
\setlength{\abovecaptionskip}{0.1 in}
\begin{figure}[t]
\centering
\begin{subfigure}[]{1.5 in}
\includegraphics[width=1.5 in,height=1.7 in]{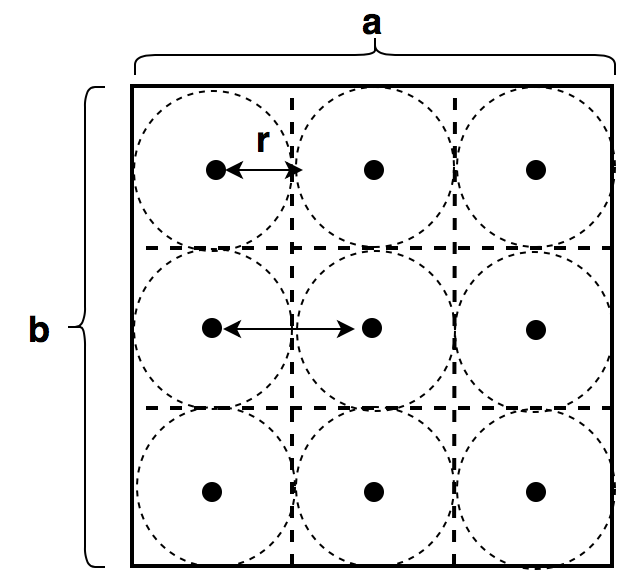}
\caption{A room with N Cells.}
\label{ExtRoom1}
\end{subfigure}
~
\begin{subfigure}[]{1.5 in}
\includegraphics[width=1.5 in, height=1.7 in]{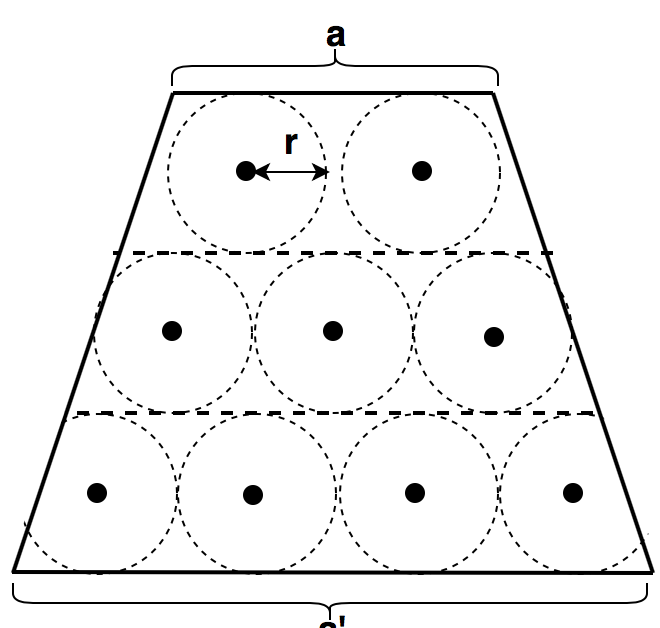}
\caption{A Hypothetical room.}
\label{ExtRoom2}
\end{subfigure}

\caption{A top view of a room with $N$ optimal THz-APs.}
\label{ExtRoom}
\end{figure}

\subsection{Extension to a General Room} \label{genroom}

Now that we have optimal number of THz-APs for the hypothetical room, we would like to convert it back to its original rectangular room of size $a=N_xg_x$ and $b=N_yg_y$. Since both the actual room and the hypothetical room have the same area, without loss of generality, we can show that the total number of cells is as shown in Equation \ref{NosAPInRoom}. 
\abovedisplayskip=0pt
\belowdisplayskip=4pt
\begin{eqnarray}
N=\frac{L}{2r}=\frac{N_xN_yg_x}{2r}=\frac{a\frac{b}{2r}}{2r}=\frac{ab}{4r^2}
\label{NosAPInRoom}
\end{eqnarray}

Combining Equations \ref{RomL} and \ref{NosAPInRoom} we can derive a relationship between the rectangular and hypothetical room, as shown in Equation \ref{RoomsRel}. 

\abovedisplayskip=-4pt
\belowdisplayskip=4pt
\begin{eqnarray}
L=\frac{ab}{2r}
\label{RoomsRel}
\end{eqnarray}

Equation \ref{RoomsRel} highlights that all rectangular rooms of same area will require the same number of optimal THz-APs, i.e., if $a_1*b_1=a_2*b_2$ then $N_1=N_2$. 

Equation \ref{RoomsRel} can be extended for any irregular room, like a trapezoid, which represents a conference or theater style room, or a polygonal room that can be converted to a rectangle, as shown in Fig. \ref{ExtRoom}. Thus a trapezoid of lengths $a_1$ \& $a'_1$ and width $b_1$ is equivalent to a rectangle of length $a_2=\frac{a+a'}{2}$ and width $b_2=b_1$. The motivational example shown in Fig. \ref{ExtRoom} proves that for a square room of $6$ units, which requires $9$ THz-APs, can perfectly fit those same number of THz-APs into a trapezoid of lengths $4$, $8$ units and width $6$ units. Please note due to our assumption of a THz-cell to be a circle, visually it seems that there are some white spaces; however, the hexagonal cell shape will perfectly fit without such white spaces. 

Since from Equation \ref{RoomsRel} we can infer there exists multiple shapes and sizes of rooms that can fit the same number of THz-APs, for the rest of the paper we refer the size of a room as $L$, which is equivalent to the hypothetical room.

%%%%%%%%%%%%%%%%%%%%%%%%%%%%%%%%%%%%%%%%%%%%%%%%%%%%%%%%%%%%%%%%%%%%%%%%%%%%%%%%%%%%%%%%%%%%%%%%%%%%%%%%%%%%%%%%%%%%%%%%%%

\subsection{Heterogeneous Room Profile}

So far, we have considered a homogeneous room, where each unit requires similar spectral efficiency. This assumption will yield high efficiency in the lower RF bands; however, in case of THz where the cell sizes are few meters in size and the demand is dependent on the user location, it will result in wastage of resources. To deploy THz-APs more efficiently, we need to understand that each square foot of the room is not the same. A regular room will have multiple pockets or hotspots, which have different users' density and demands; thus, they should require a fewer number of THz-APs. 

\setlength\belowcaptionskip{0in}
\setlength{\abovecaptionskip}{0 in}
\begin{figure}[t]
\centering
\includegraphics[width=3.5 in,height=2 in]{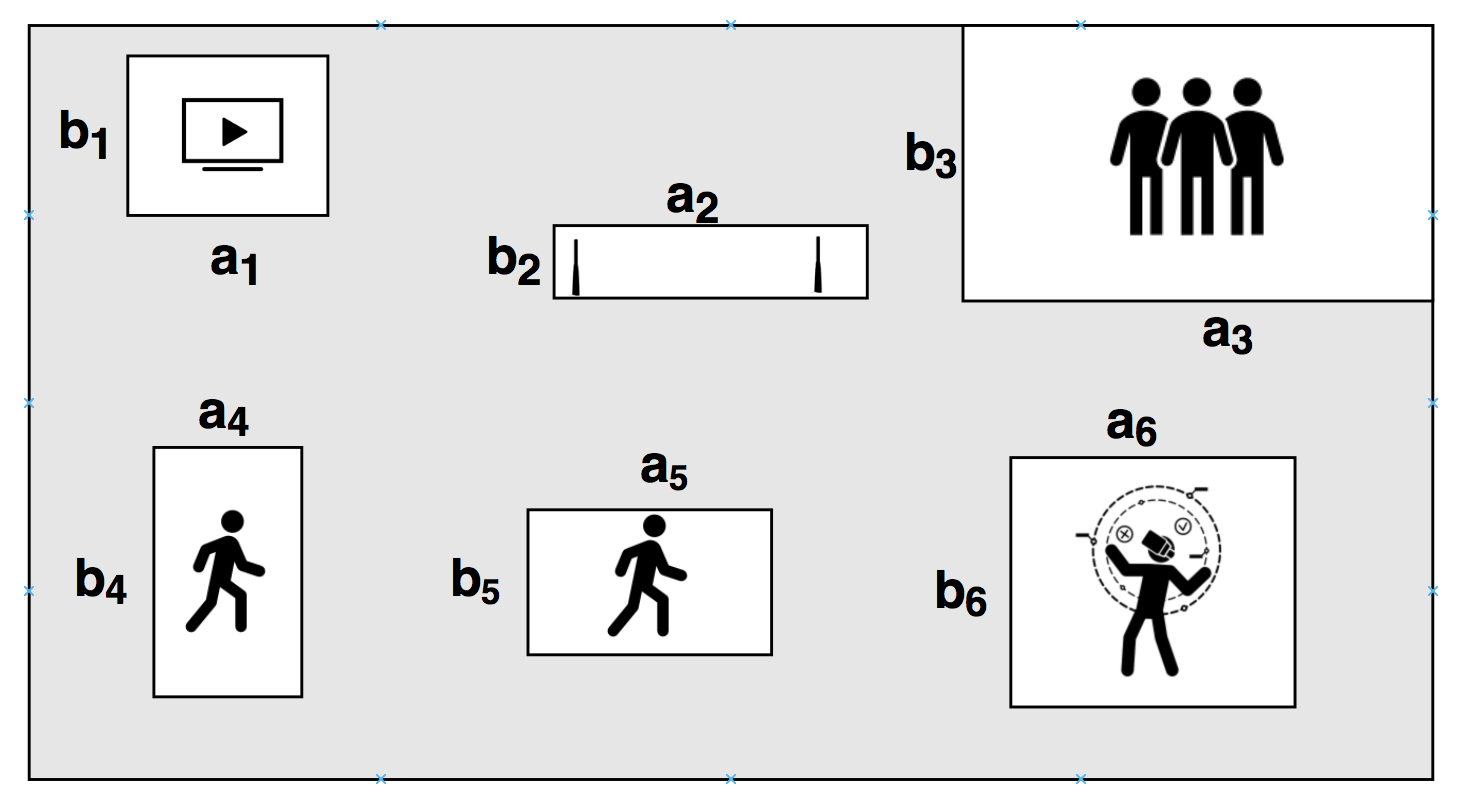}
\caption{A Heterogeneous room with multiple demand pockets.}
\label{HetRoom}
\end{figure}

Based on the type of pocket, the $i^{th}$ pocket can be bounded by a smaller rectangular sub-room of size $a_i$-by-$b_i$. Please note, for simplicity, we draw rectangles; however, the pockets can also be represented using irregular shapes and then converted into a rectangle, as explained in Section \ref{genroom}. Let us consider a room with six different types of demand pockets, as shown in Fig. \ref{HetRoom}. For static and point-to-point communication, the pocket size will be minimal and will bound the device area, as shown for pocket type $1$ and $2$. However, in case of more number of users for pocket type $3$, or users with varying mobility type on the y-axis $\Delta y$, x-axis $\Delta x$, or both axes for pocket types $4$,$5$,and $6$ respectively, the size will be proportional to the user mobility, i.e., $a_i\propto \Delta x$, $b_i\propto \Delta y$ \cite{OurGC}. By keeping the total power of the room constant, the power allocated for the $i^{th}$ pocket can be derived based on the number of users in that pocket, i.e., $\mathcal{P}_i= \mathcal{P}_o/\mathcal{J}_i$. After these pockets are identified each pocket can be considered homogeneous in itself and use Equation \ref{OptiRval} to find the optimal number of THz-APs in each pocket, $N_i=\frac{a_ib_i}{2r_i^2}$, where $r_i\propto \mathcal{P}_i=\frac{\mathcal{P}_o}{J_i}$.

%%%%%%%%%%%%%%%%%%%%%%%%%%%%%%%%%%%%%%%%%%%%%%%%%%%%%%%%%%%%%%%%%%%%%%%%%%%%%%%%%%%%%%%%%%%%%%%%%%%%%%%%%%%%%%%%%%%%%%%%%%

\setlength\belowcaptionskip{-0.1 in}
\setlength{\abovecaptionskip}{0.1 in}
\begin{figure}[t]
\centering
\begin{subfigure}[]{1.5 in}
\includegraphics[width=1.5 in,height=1.7 in]{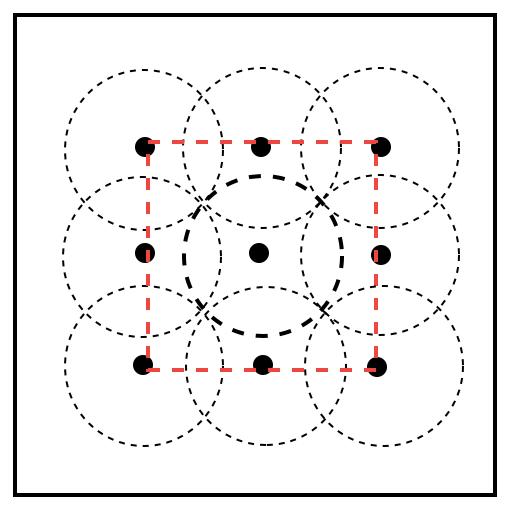}
\caption{Square Packing.}
\label{ExtRoom2}
\end{subfigure}
~
\begin{subfigure}[]{1.5 in}
\includegraphics[width=1.5 in,height=1.7 in]{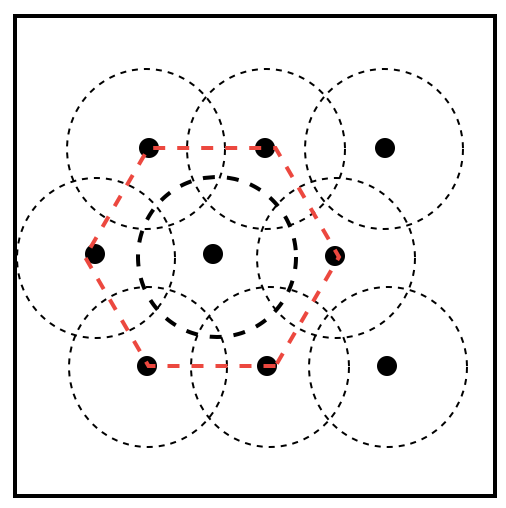}
\caption{Hexagonal Packing.}
\label{ExtRoom2}
\end{subfigure}

\caption{Cell packing schemes.}
\label{CellPack}
\end{figure}

\subsection{Optimal Room Size} 

Although we have the optimal number of THz-APs for a room of size $L$, it will also be interesting to bound the maximum room size for a fixed number of THz-APs. By simple mathematical manipulating of Equation \ref{ShanN} and combining constants, we get the Equation \ref{RM2}, where $K_2=\frac{4\mathcal{P}_o G_t (\delta)*G_r (\delta) c^2}{N*N_f*(2^\mathcal{S}-1)(4\pi f_cL)^2}$ and $\tau_2=\mathcal{K}(f_c,\rho,T)/2N$. 

\abovedisplayskip=0pt
\belowdisplayskip=4pt
\begin{eqnarray}
L^2 e^{\tau_2 L}=K_2
\label{RM2}
\end{eqnarray}

With sufficient mathematical manipulation we can convert Equation \ref{RM2} to represent the inverse Lambert function, as shown in Equation \ref{RM3}. 

\abovedisplayskip=0pt
\belowdisplayskip=4pt
\begin{eqnarray}
\begin{multlined}
L e^{\frac{\tau_2 L}{2}}=\pm\sqrt{K_2} \\ 
\frac{\tau_2L}{2} e^{\tau_2 L/2}=\pm\frac{\sqrt{\tau_2^2K_2}}{2} 
\end{multlined}
\label{RM3}
\end{eqnarray}

Now that Equation \ref{RM3} resembles the inverse Lambert function, we can solve for $L$ as shown in Equation \ref{LOpti}. There are two roots to the Equation \ref{LOpti}. Since length cannot be negative, we neglect the positive root of the function which makes $\mathcal{W}(\cdot)$ negative in Equation \ref{LOpti}.
\abovedisplayskip=0pt
\belowdisplayskip=4pt
\begin{eqnarray}
L = \ceil*{\frac{2\mathcal{W}(-\frac{\sqrt{K_2\tau_2^2}}{2})}{\tau_2}}
\label{LOpti}
\end{eqnarray}

%%%%%%%%%%%%%%%%%%%%%%%%%%%%%%%%%%%%%%%%%%%%%%%%%%%%%%%%%%%%%%%%%%%%%%%%%%%%%%%%%%%%%%%%%%%%%%%%%%%%%%%%%%%%%%%%%%%%%%%%%%
%%%%%%%%%%%%%%%%%%%%%%%%%%%%%%%%%%%%%%%%%%%%%%%%%%%%%%%%%%%%%%%%%%%%%%%%%%%%%%%%%%%%%%%%%%%%%%%%%%%%%%%%%%%%%%%%%%%%%%%%%%
%%%%%%%%%%%%%%%%%%%%%%%%%%%%%%%%%%%%%%%%%%%%%%%%%%%%%%%%%%%%% %%%%%%%%%%%%%%%%%%%%%%%%%%%%%%%%%%%%%%%%%%%%%%%%%%%%%%%%%%%%%   

\section{Efficient Cell Packing Schemes} \label{CPack}

Outages due to blockages and environmental uncertainties can significantly impact the average throughput of the THz-APs within the demand pockets. To ensure minimum outages for mobile users during handoff, the THz-APs need to overlap with each other or have redundancies spatially. Moreover, As shown in Fig. \ref{CellPack}, overlapping or packing of cells can lead to extra white spaces in the room, eventually increasing the total number of THz-APs. Even the way these THz-APs are packed in the room can affect the optimality analysis. For example, in Fig. \ref{CellPack} hexagonal packing is more efficient compared to square packing, which means hexagonal packing will require more THz-APs to offset the white space compared to square packing. In this section, we discuss two strategies of improving coverage while keeping the optimality analysis done in Section \ref{Ana} constant.

\setlength\belowcaptionskip{0 in}
\setlength{\abovecaptionskip}{0 in}
\begin{figure}[t]
\centering
\begin{subfigure}[]{1.5 in}
\includegraphics[width=1.5 in,height=1.7 in]{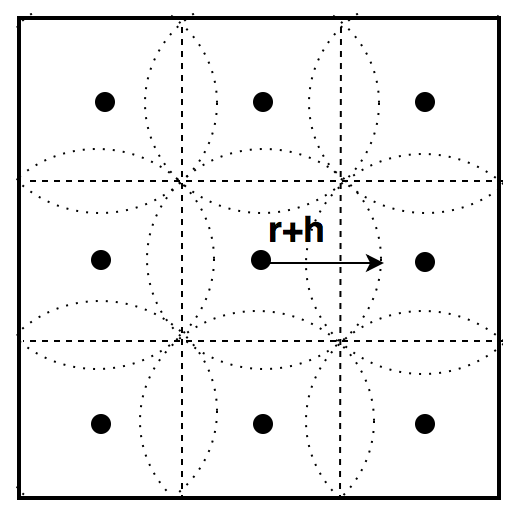}
\caption{Increase Cell Radius.}
\label{RepRoom1}
\end{subfigure}
~
\begin{subfigure}[]{1.5 in}
\includegraphics[width=1.5 in,height=1.7 in]{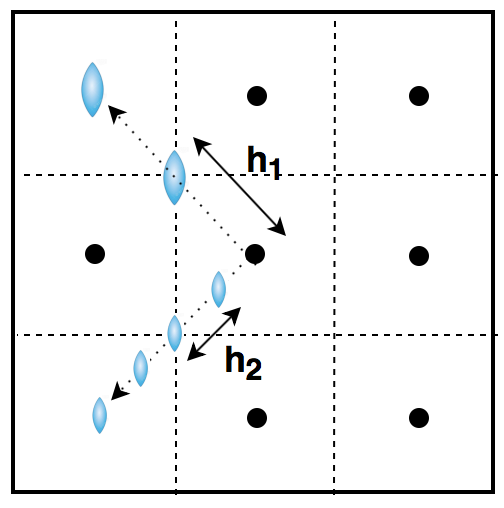}
\caption{Repeater Assisted Room.}
\label{RepRoom2}
\end{subfigure}

\caption{Schemes for efficient THz-AP packing.}
\label{RepRoom}
\end{figure}

%%%%%%%%%%%%%%%%%%%%%%%%%%%%%%%%%%%%%%%%%%%%%%%%%%%%%%%%%%%%%%%%%%%%%%%%%%%%%%%%%%%%%%%%%%%%%%%%%%%%%%%%%%%%%%%%%%%%%%%%%%

\subsection{Radius Increase} \label{Rinc}
The first solution to avoid an increase in THz-APs is to increase the radius of the cells by just $h$ units over the optimal cell size of $r$ units, as shown in Fig. \ref{RepRoom}. By keeping the number of THz-APs fixed to $N$, obtained from Equation \ref{OptiNval}, we can derive the new radius of the THz-cells that just overlaps $h$ units, as shown in Equation \ref{RI1}, where $r'=r+h$.  

\abovedisplayskip=0pt
\belowdisplayskip=4pt
\begin{eqnarray}
2^\mathcal{S}-1=\frac{\frac{\mathcal{P}_o}{N}G_t (\delta)G_r (\delta)}{N_D(\frac{4\pi f_cr'}{c})^2e^{\mathcal{K}r'}}
\label{RI1}
\end{eqnarray}

Using mathematical manipulation and Lambert function we can show that it represents the same form as Equation \ref{RM2}. Thus, solving for $r'$ (using similar manipulation done for Equation \ref{RM3}), as shown in Equation \ref{RI2}, where $K_3=\frac{4\mathcal{P}_o G_t (\delta)*G_r (\delta) c^2}{N*N_f*(2^\mathcal{S}-1)(4\pi f_c)^2}$ and $\tau_3=\mathcal{K}(f_c,\rho,T)$, and later deriving $h$, as shown in Equation \ref{RI3}. If we keep operating frequency constant, the factor that can offset the $h$ increment is the $K_3$ parameter which is $\propto G_t (\delta)G_r (\delta)$. We discuss the merits of this scheme in Section \ref{NuAna}.

\abovedisplayskip=0pt
\belowdisplayskip=4pt
\begin{eqnarray}
r'^2 e^{\tau_3 r'}=K_3
\label{RI2}
\end{eqnarray}

or, 

\abovedisplayskip=0pt
\belowdisplayskip=4pt
\begin{eqnarray}
h= \ceil*{\frac{2\mathcal{W}(-\frac{\sqrt{K_3\tau_3^2}}{2})}{\tau_3}}-r
\label{RI3}
\end{eqnarray}

%%%%%%%%%%%%%%%%%%%%%%%%%%%%%%%%%%%%%%%%%%%%%%%%%%%%%%%%%%%%%%%%%%%%%%%%%%%%%%%%%%%%%%%%%%%%%%%%%%%%%%%%%%%%%%%%%%%%%%%%%

\subsection{Repeater Assistance}

There are multiple caveats to the Radius Increase scheme, explained later in Section \ref{NuAna}, which can be a nonviable solution for some parts of the THz spectrum (especially the upper end of the spectrum). Given our power and grid distribution constraints, these THz-cells can only operate till a distance $r$, without any radius increase scheme, as explained in Section \ref{Rinc}, and the distance between two cells is $2r$. Thus, an alternative solution is to use repeaters, such as lenses, mirrors, reflectarrays, or hyper-surfaces \cite{ComDP}, to reduce the distance between two THz-cells, eventually merging these two adjacent cells. As shown in Fig. \ref{RepRoom}(b), by using repeaters or relays, we can extend the coverage of a single cell to $2r$. However, the feasibility is dependent on the number of repeaters required for this cell merger strategy. As shown in Fig. \ref{RepRoom}(b), some THz frequencies might require two repeaters, while some might need an array of repeaters.

In this scheme, we replace the neighboring THz-AP, which is $2r$ distance away, with $m$ number of repeaters along that path. Therefore, the new worst-case distance between two hops is $r'=2r/m$, where $m$ is the number of repeaters. We assume that each repeater is equipped with a signal amplification, which can offset the attenuation of the signal during each hop. Moreover, when we combine two THz-APs, we will also combine the transmit power to $\frac{2\mathcal{P}_o}{N}$ and the required spectral efficiency to $2\mathcal{S}$. Thus, for a fixed number of THz-APs $N$, obtained from Equation \ref{OptiNval}, we can rewrite Equation \ref{ShanN} to Equation \ref{ReA1}.

\abovedisplayskip=0pt
\belowdisplayskip=4pt
\begin{eqnarray}
2^{2\mathcal{S}}-1=\frac{\frac{2\mathcal{P}_o}{N}G_t (\delta)G_r (\delta)}{N_D(\frac{4\pi f_c}{c})^2(r')^2e^{\mathcal{K}r'}}
\label{ReA1}
\end{eqnarray}

\abovedisplayskip=0pt
\belowdisplayskip=4pt
\begin{eqnarray}
 1/m e^{\tau_4/m}=K_4
\label{ReA2}
\end{eqnarray}

Using mathematical manipulation on Equation \ref{ReA2} we can solve for $m$, the number of repeaters, as shown in Equation \ref{ReA2}, where $K_4=\frac{2\mathcal{P}_o G_t (\delta)*G_r (\delta) c^2}{N*N_f*(2^(2\mathcal{S})-1)(4\pi f_c)^2(2r)^2}$ and $\tau_4=2r\mathcal{K}(f_c,\rho,T)$. 

\abovedisplayskip=0pt
\belowdisplayskip=4pt
\begin{eqnarray}
m= \ceil*{\frac{\tau_4}{\mathcal{W}(\tau_4 K_4)}}
\label{ReA3}
\end{eqnarray}

%%%%%%%%%%%%%%%%%%%%%%%%%%%%%%%%%%%%%%%%%%%%%%%%%%%%%%%%%%%%%%%%%%%%%%%%%%%%%%%%%%%%%%%%%%%%%%%%%%%%%%%%%%%%%%%%%%%%%%%%%%
%%%%%%%%%%%%%%%%%%%%%%%%%%%%%%%%%%%%%%%%%%%%%%%%%%%%%%%%%%%%%%%%%%%%%%%%%%%%%%%%%%%%%%%%%%%%%%%%%%%%%%%%%%%%%%%%%%%%%%%%%%
%%%%%%%%%%%%%%%%%%%%%%%%%%%%%%%%%%%%%%%%%%%%%%%%%%%%%%%%%%%%%%%%%%%%%%%%%%%%%%%%%%%%%%%%%%%%%%%%%%%%%%%%%%%%%%%%%%%%%%%%%%  
 
\setlength\belowcaptionskip{-0.1in}
\setlength{\abovecaptionskip}{0in}
\begin{figure}[t]
\centering
\includegraphics[width=3.5 in,height=2 in]{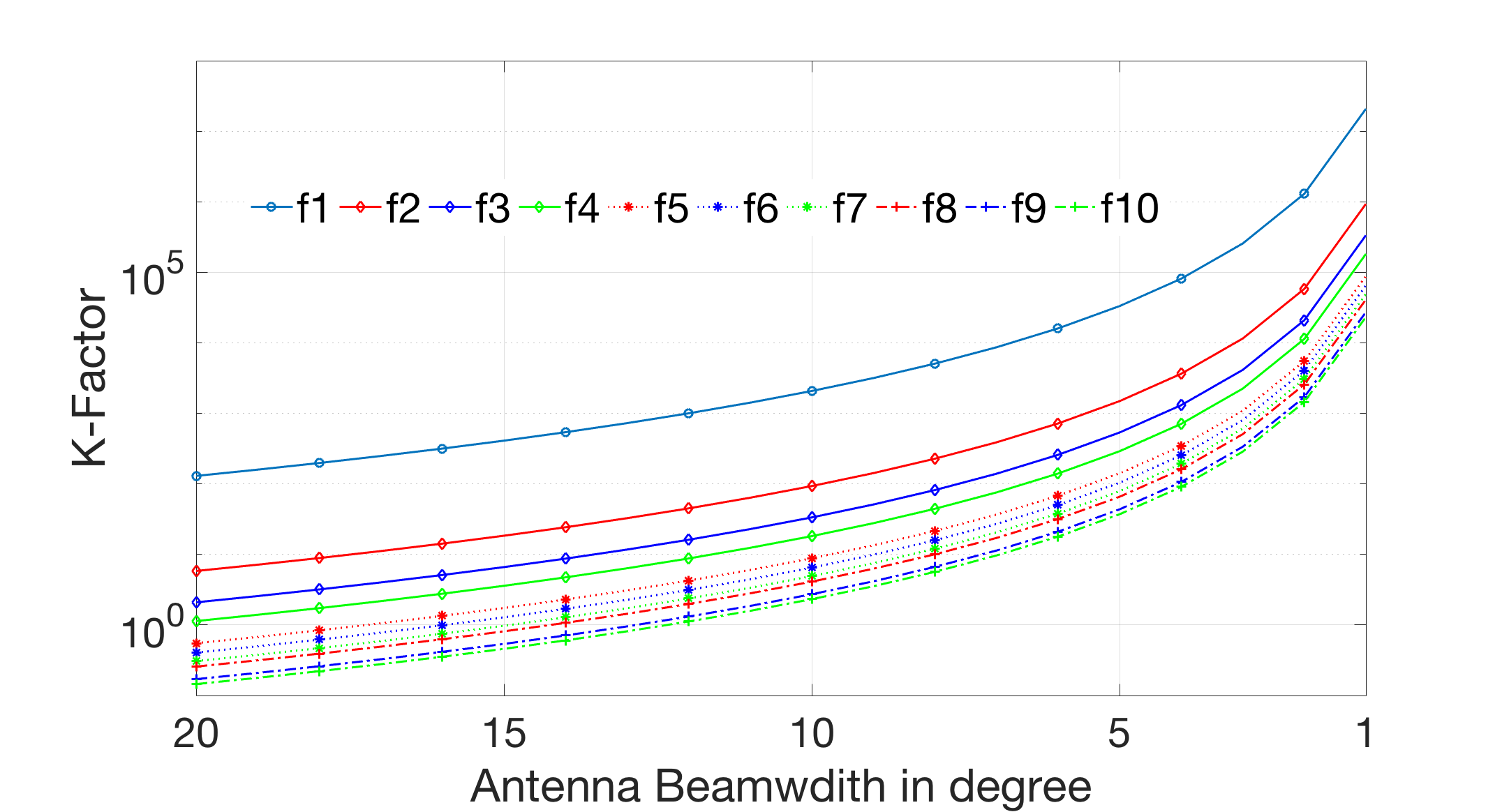}
\caption{K-Factor analysis for different Antenna Beamwidth.}
\label{KPlot}
\end{figure}

\setlength\belowcaptionskip{-0.1in}
\setlength{\abovecaptionskip}{0 in}
\begin{figure}[t]
\centering
\includegraphics[width=3.5 in,height=2 in]{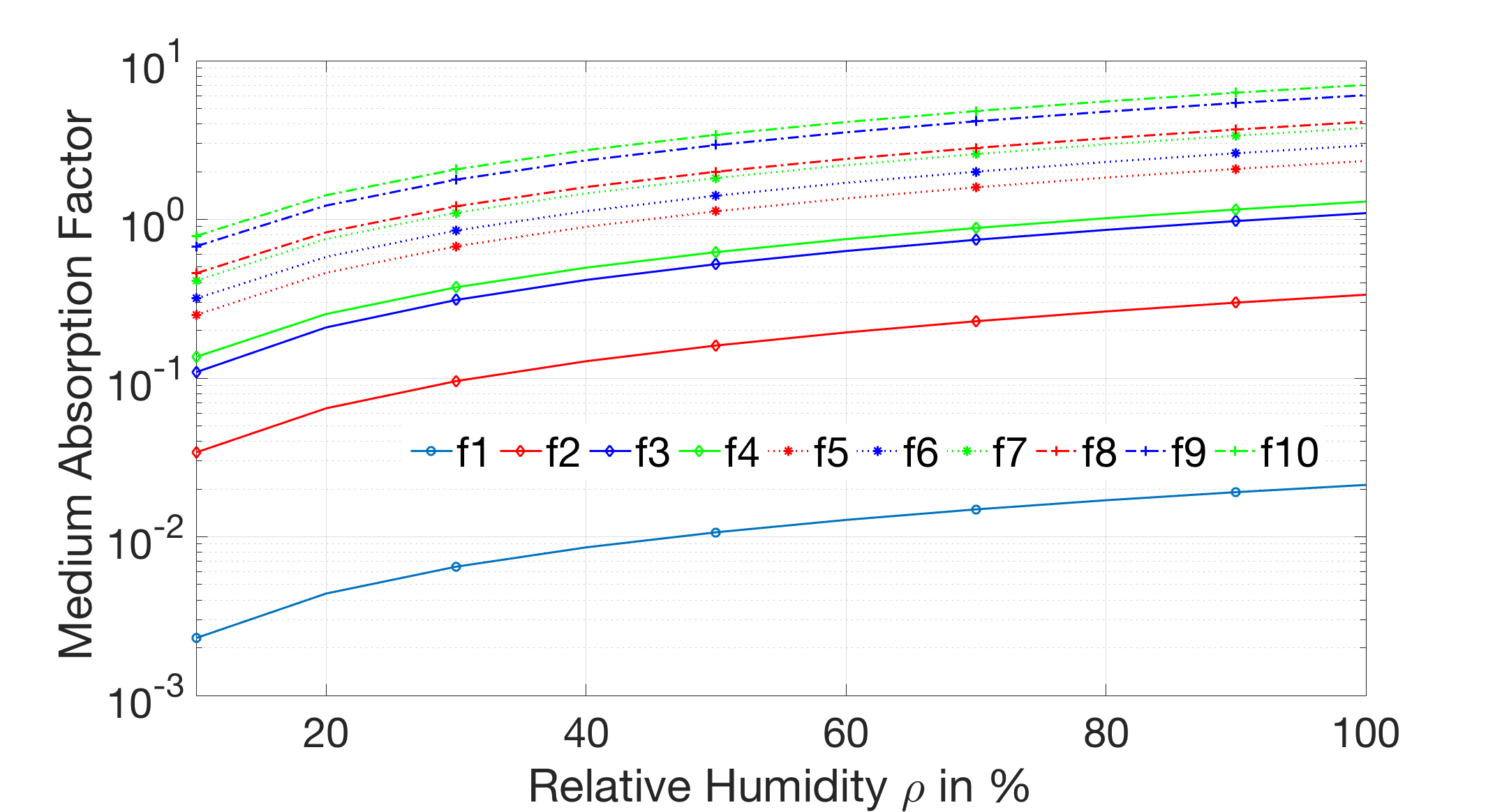}
\caption{$\tau$ factor analysis for different Relative Humidity $\rho$.}
\label{TauPlot}
\end{figure}

\section{Numerical Analysis} \label{NuAna}

In this section, we perform an extensive analysis for the analytical models presented throughout the paper. 

%%%%%%%%%%%%%%%%%%%%%%%%%%%%%%%%%%%%%%%%%%%%%%%%%%%%%%%%%%%%%%%%%%%%%%%%%%%%%%%%%%%%%%%%%%%%%%%%%%%%%%%%%%%%%%%%%%%%%%%%%%

\subsection{Simulation Parameters}

The optimal value analysis shown in Sections \ref{Ana} and \ref{CPack} are dependent on a set of environmental and hardware factors. For our numerical analysis, we used the parameters mentioned in Table \ref{ParaTab}, otherwise discussed separately in the text. Although the optimal values can be modeled for any type of room, in this paper, we limit our analysis to homogeneous user distribution, i.e., each cell requests the same spectral efficiency $\mathcal{S}$. We assume an average relative humidity of $60\%$ at a room temperature of $25\degree C$. Although the sub-channels introduced in Table \ref{Tab1} have an abundance of contiguous bandwidth, to ensure spectral efficiency, we constrain the usable bandwidth to $10GHz$. THz spectrum will require very narrow beamwidth to offset the huge path losses. However, it was shown in \cite{OurGC} that there exists optimal antenna beamwidth based on user mobility, after which the average throughput will decline. We start our analysis with $\delta=20\degree$ and then observe the change in optimal values for narrower antenna beamwidths \cite{OurGC}. 

\setlength{\intextsep}{0pt}% Remove \textfloatsep
\setlength\belowcaptionskip{0 in}
\setlength\abovecaptionskip{0 in}
\begin{table}[t]
\centering
\caption{Simulation Parameters}
\begin{tabular}{ |c|c|c| } 
 \hline
 Parameter & Description & Value \\ \hline \hline
 $\rho$ & Relative Humidity & $60\%$ \\ \hline
 $T$ & Room Temperature & $25\degree C$ \\ \hline
 $\mathcal{P}_o$ & Room Transmit Power & $0dBm$ \\ \hline
 $N_f$ & Noise Spectral Density& $-193dB/GHz$\\ \hline
 $\mathcal{R}$ & Requested Data Rate  & $1Gbps$ \\ \hline
 $\mathcal{B}$ &Available Bandwidth & $10Ghz$\\ \hline
 $\mathcal{S}$ &Spectral Efficiency & $0.1Gbps/GHz$\\ \hline
 $\delta$ & Antenna Beamwidth & $20\degree$ \\ \hline
 $L$ & Hypothetical Room Length & $10m$ \\ \hline
 $r$ & THz-Cell Radius & $0.56m$ ($N=9$) \\ \hline
\end{tabular}
\label{ParaTab}
\end{table}

\setlength\belowcaptionskip{-0.1in}
\setlength{\abovecaptionskip}{0 in}
\begin{figure}[t]
\centering
\includegraphics[width=3.5 in,height=2 in]{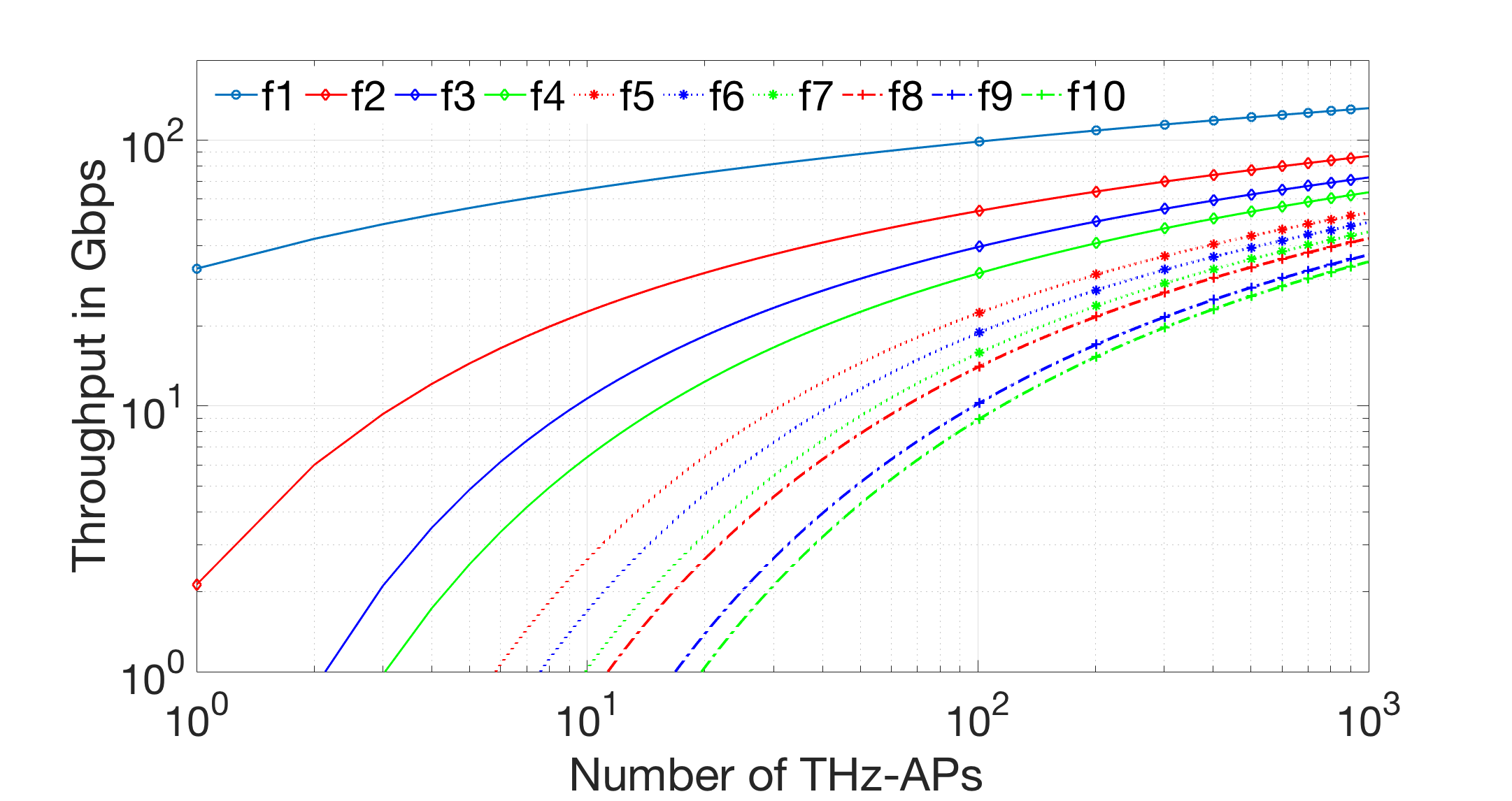}
\caption{Optimal Number of THz-APs required in a room described in Table \ref{ParaTab}.}
\label{NOpplot}
\end{figure}

\setlength\belowcaptionskip{-0.1in}
\setlength{\abovecaptionskip}{0 in}
\begin{figure}[t]
\centering
\includegraphics[width=3.5 in,height=2 in]{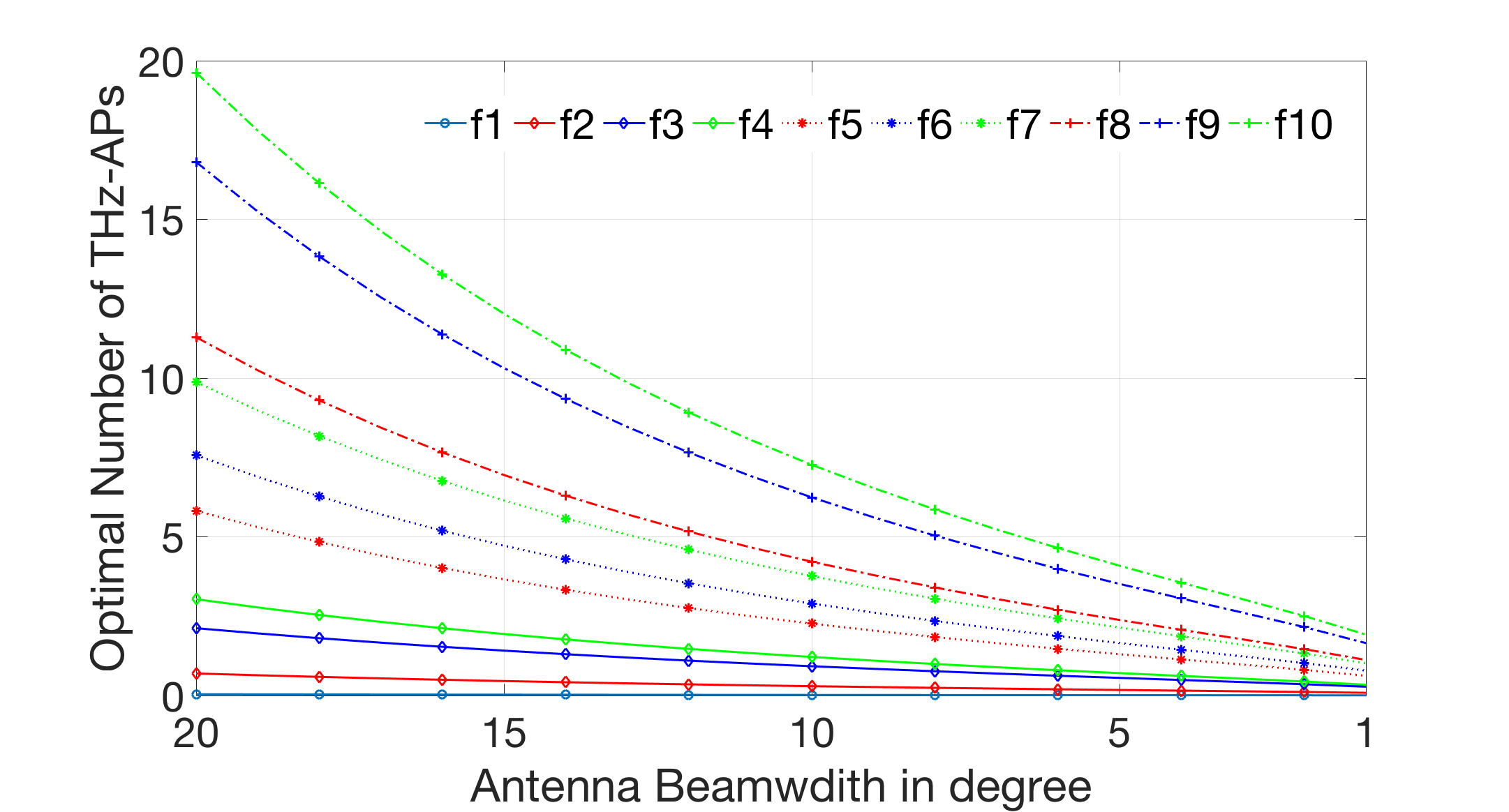}
\caption{Optimal number of THz-APs for varying Antenna Gain to achieve 0.1Gbps/GHz.}
\label{NOpGplot}
\end{figure}

\setlength{\intextsep}{0pt}% Remove \textfloatsep
\setlength\belowcaptionskip{9pt}
\setlength\abovecaptionskip{0pt}
\begin{table*}[t]  
\caption{Optimal Number of THz-AP required for varying Spectral Efficiency $\mathcal{S}$ in a room described in Table \ref{ParaTab}}
\centering
\begin{tabular}{ |p{2cm}|p{1cm}|p{1cm}|p{1cm}|p{1cm}|p{1cm}|p{1cm}|p{1cm}|p{1cm}|p{1cm}|p{1cm}|} 
\hline
$\mathcal{S}$		& $f_1$ & $f_2$ & $f_3$ & $f_4$ & $f_5$ & $f_6$ & $f_7$ & $f_8$ & $f_9$ & $f_{10}$ \\ \hline
$2Gbps/GHz$		& $1$ & $	9$ & $23$ & $41$ & $83$ & $112$ & $147$ & $175$ & $263$ & $ 308$ \\ \hline
$1Gbps/GHz$		& $1$ & $	4$ & $10$ & $16$ & $32$ & $42$ & $56$ & $66$ & $98$ & $115$ \\ \hline
$0.5Gbps/GHz$	& $1$ & $	2$ & $6$ & $9$ & $16$ & $22$ & $28$ & $33$ & $49$ & $58$ \\ \hline
$0.1Gbps/GHz$	& $1$ & $	1$ & $3$ & $4$ & $6	$ & $8$ & $10$ & $12$ & $17$ & $20$ \\ \hline	     
\end{tabular}
\label{NOpTab}
\vspace{-4mm}
\end{table*}

%%%%%%%%%%%%%%%%%%%%%%%%%%%%%%%%%%%%%%%%%%%%%%%%%%%%%%%%%%%%%%%%%%%%%%%%%%%%%%%%%%%%%%%%%%%%%%%%%%%%%%%%%%%%%%%%%%%%%%%%%%
\subsection{Evaluation of $K$ \& $\tau$}
Before we compute the optimal values for a room, we need to explore more about the factors $K$ and $\tau$, which exists in different forms throughout the analysis. If we keep frequency and other environmental parameters (like, room dimension, and relative humidity) constant, the variables $K_1$ through $K_4$ are predominantly dependent on the antenna gain (or the antenna beamwidth $\delta$). As shown in Fig. \ref{KPlot}, the K-Factor value increases exponentially as we move closer to $\delta=1\degree$, which will result in fewer THz-APs. The $K$-factor improvement is significant for lower frequency sub-channels, while the higher frequency sub-channels are indifferent to the beamwidth improvement. On the other hand, variables $\tau_1$ through the $\tau_4$ are dependent on medium absorption coefficient $\mathcal{K}(f_c,\rho,T)$. Fig. \ref{TauPlot} shows that the coefficient increases with increasing relative humidity, which results in higher path loss. 

%%%%%%%%%%%%%%%%%%%%%%%%%%%%%%%%%%%%%%%%%%%%%%%%%%%%%%%%%%%%%%%%%%%%%%%%%%%%%%%%%%%%%%%%%%%%%%%%%%%%%%%%%%%%%%%%%%%%%%%%%%

\subsection{Optimal Number of APs}
By using the parameter values in Table \ref{ParaTab} with Equation \ref{OptiNval}, we can find the optimal number of THz-APs required for a room of $L=10m$ at $\delta=20\degree$. Fig. \ref{NOpplot} shows the optimal number of APs for different frequencies. Even with $1000$ Thz-APs, none of the sub-channels except $f_1$ can achieve $100Gbps$ throughput. Moreover, due to the power constraint throughput for all of the sub-channels converge as the number of APs increase. If we compare these sub-channels on the spectral efficiency scale, shown in Table \ref{NOpTab}, the higher sub-channels can reach $100Gbps$ for a much lower number of THz-APs if we use $100$s of $GHz$ of contiguous bandwidth. If we want to use higher frequency sub-channels for high mobile users (i.e., pocket type $6$, shown in Fig. \ref{HetRoom}) we will have to use large amount of bandwidth. On the contrary, for users with static or nomadic applications, we can use narrower beamwidth and reduce the number of THz-APs significantly, as shown in Fig. \ref{NOpGplot}. The higher sub-channels require $<10$ THz-APs to achieve a spectral efficiency of $0.1Gbps/GHz$ at $\delta=10\degree$.

%%%%%%%%%%%%%%%%%%%%%%%%%%%%%%%%%%%%%%%%%%%%%%%%%%%%%%%%%%%%%%%%%%%%%%%%%%%%%%%%%%%%%%%%%%%%%%%%%%%%%%%%%%%%%%%%%%%%%%%%%%

\subsection{Optimal Room Size}
\begin{figure}[t]
\centering
\includegraphics[width=3.5 in,height=2 in]{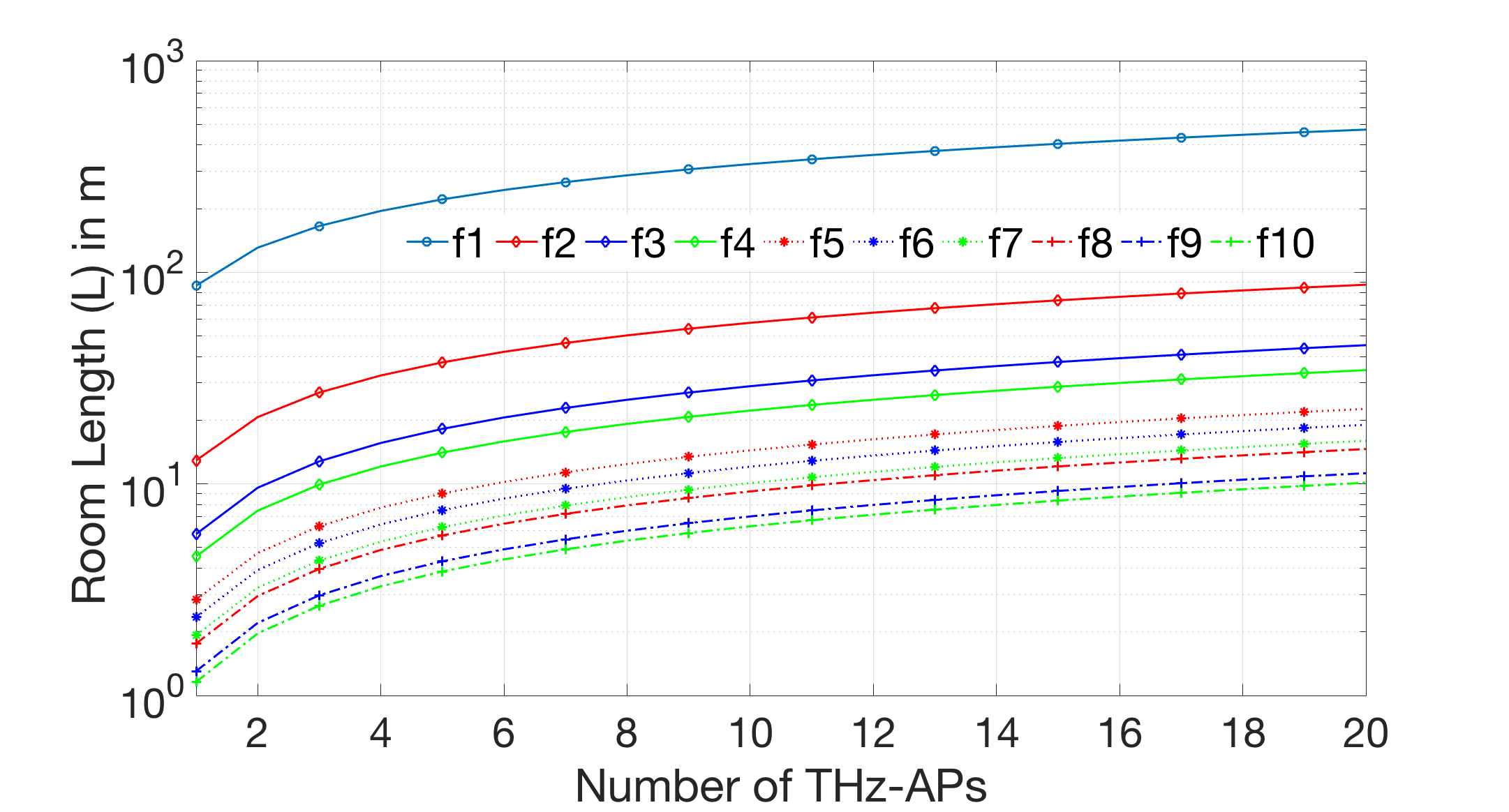}
\caption{Optimal Room Size for a fixed number of THz-APs.}
\label{RLplot}
\end{figure}

Fig. \ref{RLplot} shows the optimal room length $L$ for a fixed number of THz-APs. It seems that with a limited number of THz-APs, it is hard to achieve even a minimum spectral efficiency of $0.1Gbps/GHz$. The optimal room lengths are very small for higher sub-channels. This relationship will help define the size of the demand pockets that can fit with a fixed amount of THz-APs in a room. For example, at higher sub-channels and relatively lower number of THz-APs, it is likely the pocket sizes will be smaller, resembling pocket types $1$ and $2$, as shown in Fig \ref{HetRoom}. While for larger pocket sizes, the sub-channels $f_1$ through $f_3$ can efficiently be deployed with just $\le4$ APs.

\setlength\belowcaptionskip{-0.1in}
\setlength{\abovecaptionskip}{0 in}

%%%%%%%%%%%%%%%%%%%%%%%%%%%%%%%%%%%%%%%%%%%%%%%%%%%%%%%%%%%%%%%%%%%%%%%%%%%%%%%%%%%%%%%%%%%%%%%%%%%%%%%%%%%%%%%%%%%%%%%%%%
\subsection{Radius Increase}

\begin{figure}[t]
\centering
\includegraphics[width=3.5 in,height=2 in]{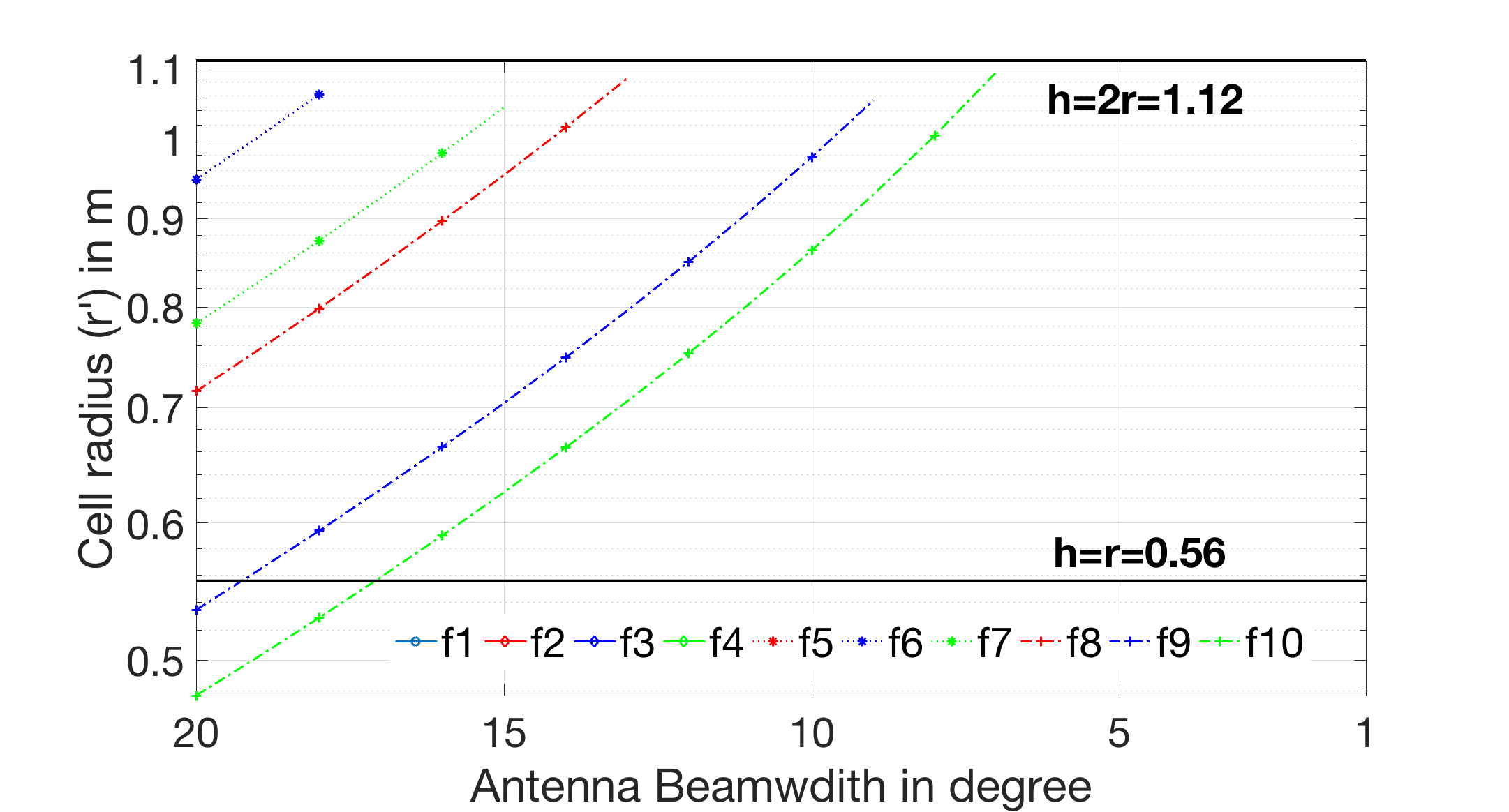}
\caption{Overlap Factor analysis for different Antenna Beamwidth for N=9.}
\label{Overplot}
\end{figure}

When we try to increase the radius of a THz-AP while keeping the frequency constant, we are effectively trying to figure out the improvement required in factor $K_3$, which is $\propto G_t (\delta)*G_r (\delta)$. By improving the antenna gains (i.e., narrower beamwidths), we can effectively pack the cells while keeping the number of THz-APs constant. Although narrower antenna beamwidth can improve cell-packing, there is a limitation to this scheme. Firstly, antenna gains are dependent on the antenna beamwidth $\delta$; due to lack of sophisticated hardware in the THz, there is a physical limitation to the width of the beam. Secondly, as we keep on narrowing the beam, the beams are more susceptible to mobility induced outages \cite{OurCCNC1}. Even small-scale mobility, such as head and body movement, can impact the connectivity \cite{OurGC}.Thus there is both a physical and practical upper bound to the antenna gain we can achieve. Let us consider a fixed number of THz-APs $N=9$, as shown in the motivational example in Fig. \ref{ExtRoom}. The cell radius for varying antenna beamwidth is shown in Fig. \ref{Overplot}, which is clipped off between at $r'=2r$ and $r'=3r$, i.e., $h=r$ and $h=2r$ from Equation \ref{RI3}. Sub-channels beyond $f_6$ show up in the clipped off version of Fig. \ref{Overplot}, since the $9$ THz-APs, is more than what is required for the room, as shown in Table \ref{NOpTab}.

\setlength\belowcaptionskip{-0.1in}
\setlength{\abovecaptionskip}{0 in}

%%%%%%%%%%%%%%%%%%%%%%%%%%%%%%%%%%%%%%%%%%%%%%%%%%%%%%%%%%%%%%%%%%%%%%%%%%%%%%%%%%%%%%%%%%%%%%%%%%%%%%%%%%%%%%%%%%%%%%%%%%
\subsection{Number of Repeaters}

\begin{figure}[t]
\centering
\includegraphics[width=3.5 in,height=2 in]{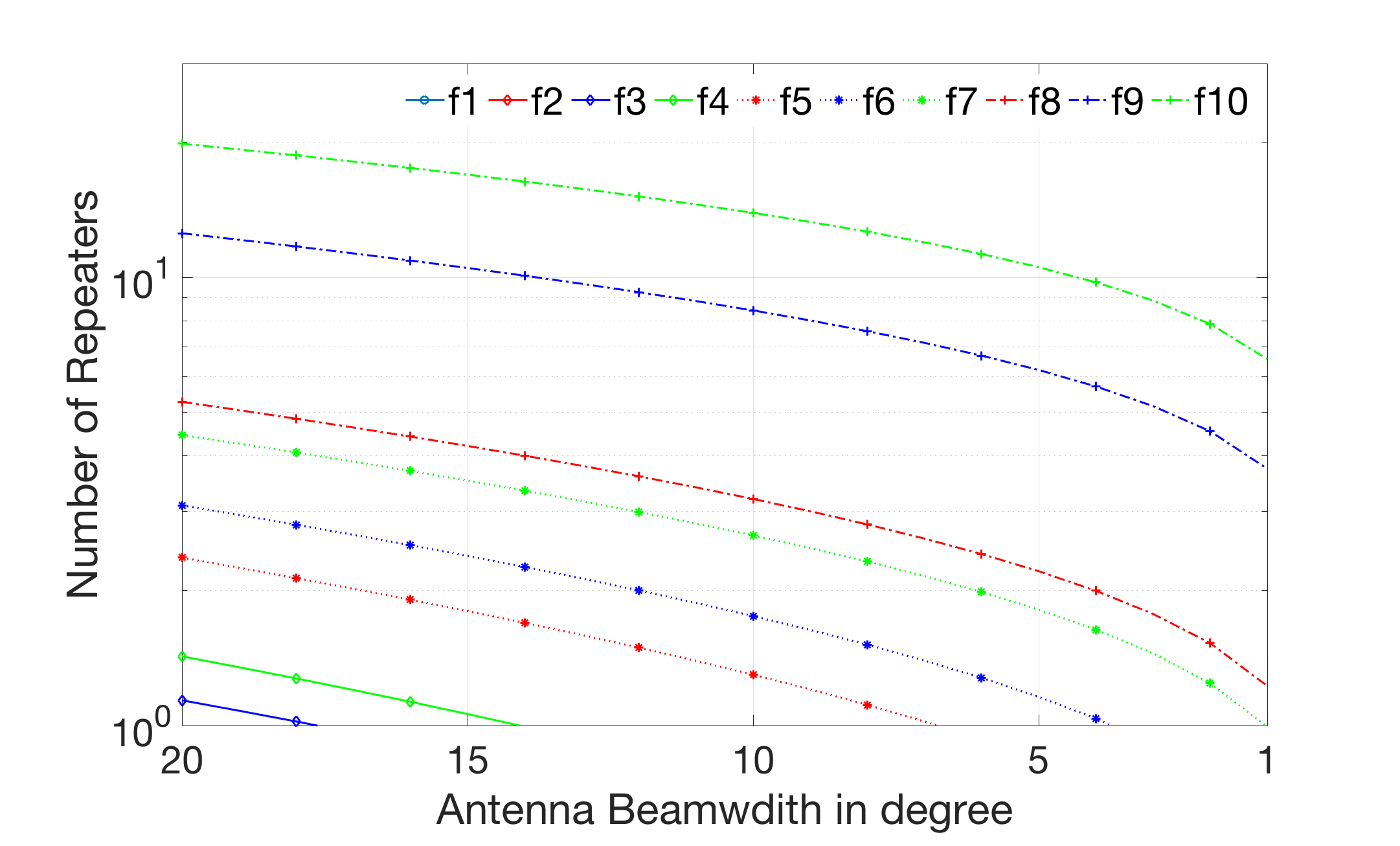}
\caption{Optimal Repeaters required to combine 2 THz-APs for N=9.}
\label{Pepplot}
\end{figure}

The radius increase strategy is a nonviable solution for mobile users in the higher sub-channels of the THz spectrum. Therefore, using repeaters and merging two adjacent THz-APs might be a viable solution for high speed mobility in higher sub-channels. Compared to Radius Increase method, which impact all adjacent neighbors, Repeater Assistance is targeted for a single neighboring cell. The number of repeaters required for varying antenna gain is shown in Fig. \ref{Pepplot}. In the case of the example shown in Fig. \ref{ExtRoom}, the lower sub-channels $f_1$ and $f_2$ do not require any repeaters for cell merger. While for antenna beamwidth $\delta=20\degree$, the higher sub-channels will require multiple repeaters to merge two THz-cells. A combination of narrower antenna beamwidth and few repeaters might be a viable solution for mobile use at higher sub-channels.

%Missing frequencies... since these were fractional THz-APs, it is equivalent to using a fraction of the power... the fraction power not considered can be used to improve coverage....
%%%%%%%%%%%%%%%%%%%%%%%%%%%%%%%%%%%%%%%%%%%%%%%%%%%%%%%%%%%%%%%%%%%%%%%%%%%%%%%%%%%%%%%%%%%%%%%%%%%%%%%%%%%%%%%%%%%%%%%%%%
%\subsection{Usable THz Frequency}

%%%%%%%%%%%%%%%%%%%%%%%%%%%%%%%%%%%%%%%%%%%%%%%%%%%%%%%%%%%%%%%%%%%%%%%%%%%%%%%%%%%%%%%%%%%%%%%%%%%%%%%%%%%%%%%%%%%%%%%%%%
%%%%%%%%%%%%%%%%%%%%%%%%%%%%%%%%%%%%%%%%%%%%%%%%%%%%%%%%%%%%%%%%%%%%%%%%%%%%%%%%%%%%%%%%%%%%%%%%%%%%%%%%%%%%%%%%%%%%%%%%%%
%%%%%%%%%%%%%%%%%%%%%%%%%%%%%%%%%%%%%%%%%%%%%%%%%%%%%%%%%%%%%%%%%%%%%%%%%%%%%%%%%%%%%%%%%%%%%%%%%%%%%%%%%%%%%%%%%%%%%%%%%%

\section{Conclusion} \label{Con}

Given multiple challenges, indoor THz-APs are likely to be deployed more compactly with short-range and low-power transmitters. Narrower antenna beamwidth might help solve the issue; however, it will open up other problems such as small-scale mobility outages and the need for fast beam alignment strategy. In this paper, we present an analytical model to quantify: (a) the optimal number of THz-APs required for both homogeneous and heterogeneous user deployment, (b) optimal room length that can operate under a fixed number of THz-APs, (c) optimal cell radius increase required to improve cell packing, and (d) optimal number of the repeaters needed to merge two THz-APs. Our numerical analysis suggests that the higher sub-channels will require a significant amount of THz-APs to provide higher throughput compared to lower sub-channels. If we want to reduce the total number of THz-APs and still guarantee reliable communication, we might either need to improve gain or use repeaters or both. All of these methods will either increase outage time or increase deployment costs.The higher sub-channels are likely to be allocated for the static and nomadic indoor use-case, while the lower sub-channels are for relatively higher mobile use-cases. In the future, we would like to explore methods through which a room can be divided into smaller, more efficient demand pockets. Furthermore, special strategies can be implemented, specific to these pockets, to mitigate outages due to blockages and environmental uncertainties. 

%The lower THz spectrum till $5THz$ can be deployed at wider beamwidth with less number of APs...While if we want the same service at the higher THz, we would require more than 100 APs, or nearly 20 repeaters..

%Additionally the Placement is also a big factor 
%Number of APs also increase the number of handoffs... smart handoff strategies for ultra small cells...
%Need support from other spectrum...cross layer architecture

\setlength\belowcaptionskip{-0.1in}
\setlength{\abovecaptionskip}{0 in}

\end{document}